\documentclass[aps,10pt,twocolumn]{revtex4-1}
\usepackage{hyperref}
\usepackage{enumerate}
\usepackage{amsmath}
\usepackage{graphicx}
\usepackage{epstopdf}
\usepackage{dcolumn}%
\usepackage{bm}%
\usepackage{color}
\usepackage{subfigure}
\begin{document}

\title{ Isothermal Equation of State of Three Dimensional Yukawa Gas  } 
\author{Manish K. Shukla}
\email[shuklamanish786@gmail.com]\
\author{K. Avinash}
\affiliation{Department of Physics and Astrophysics, University of Delhi, Delhi, 110007, India}
\author{Rupak Mukherjee}
\author{R. Ganesh}
\affiliation{Institute for Plasma Research, HBNI, Gandhinagar, Gujarat, 382428, India}

% \date{\today}

\begin{abstract}  

Molecular Dynamics (MD) simulation is carried out to examine the effect of particle confinement on the pressure of 3D  Yukawa gas. Confinement effects are taken into account by using perfectly reflecting boundary condition in MD simulations. An equation of state relating pressure to number density is obtained. The results of the MD simulations show that in weak coupling regime pressure of confined Yukawa gas is much bigger than the kinetic pressure and scales quadratically with number density. Results are compared with earlier theories and experiments which show quadratic scaling of dust pressure with density.
\end{abstract}

\keywords{Yukawa gas, equation of state, dusty plasmas, MD simulation}
\pacs{51.30.+i, 52.25.Kn, 52.27.Lw, 52.65.Yy}
\maketitle 

\section{Introduction}
Complex/Dusty plasma is a system of micron sized solid/dust grains embedded in the background plasma of electrons and ions. 
The typical size of these dust grains ranges from nanometers to micrometers and mass ranges from $10^{10}-10^{12} $ times the mass of proton. In nature, dust is found almost everywhere in space such as interstellar clouds, interplanetary space, planetary rings and solar system \cite{frank2000waves}. Dusty plasma can also be formed in laboratories for example in DC and RF discharges. In laboratory, dust grains acquire a net negative charge due to higher mobility of electron than that of ions \cite{shukla2009colloquium}. For a typical micron sized dust grain, the charge is of the order of $10^{3}-10^{4} $ times the electronic charge. The Coulombic interaction between dust-dust grain gets screened in presence of the background electron-ion plasma and the interaction becomes Yukawa (i.e. screened Coulomb) potential. 
Yukawa potential has the following form, 
$$\phi(r) = \frac{Q_d}{4\pi\epsilon_0} \frac{\exp(-r/\lambda_D)}{r} $$ 
where $Q_d$ is dust charge, $r$ is the distance between two dust particles and $\lambda_D$ is the screening length. Screening length depends on density and temperature of the background plasma as $ 1/\lambda_D^2=\left(\frac{e^2n_{e}}{\epsilon_0k_BT_e}+ \frac{q^2 n_{i}}{\epsilon_0T_i}\right)$ where $n_e(n_i)$ is electron(ion) density, $e(q)$ is electron(ion) charge and $T_e (T_i)$ is electron(ion) temperature. In experimental conditions dust-neutral collisions, ion-drag etc. also affect the dust dynamics. However, the dominant interaction among particles in dusty plasmas is the Yukawa potential. The high mass and high charge are the two features of dusty plasma which distinguishes it from normal electron ion plasma. The high mass of dust grains makes the dust dynamics very slow hence, many interesting phenomena can be seen easily by simple optical instruments and sometime with naked eyes, too. Further, due to the high dust charge, whenever the electrostatic potential energy of dust-dust interaction exceeds by its thermal energy, the system undergoes a phase change from gaseous to a more ordered liquid or solid state. 

The thermodynamical equilibrium  state of a Yukawa system  may be characterized by two dimensionless parameters\cite{hamaguchi1994thermodynamics}: $\kappa=a/\lambda_D $ ; the ratio of the mean inter-particle distance $ a=\left( 3/4\pi n_d\right)^{1/3} $ (here $n_d$ is the dust number density) to the screening length $\lambda_D$ and $\Gamma={Q_d^2}/{4\pi\epsilon_0 a T_d} $ ; the inverse of dust temperature $T_d$ measured in units of $ {Q_d^2}/{4\pi\epsilon_0 a}$. The coupling parameter $\Gamma^{*}= \Gamma \exp(-\kappa)$ which is the ratio of the mean inter-particle potential energy to the mean kinetic energy, is used as a measure of coupling strength in dusty plasmas. For $\Gamma^* \ll 1$ the correlation effects are almost negligible and Yukawa system behaves like an ideal gas. For $\Gamma^* \sim 1$, system behaves like an interacting Yukawa fluid and $\Gamma^* \gg 1$ corresponds to a condensed solid state where particles arrange themselves in a regular lattice form. Thus by tuning the parameters $ \Gamma $ and $\kappa$, a variety of states of Yukawa system can be realized. 
  
Dusty plasma supports a variety of collective modes most of which are mediated through dust pressure. Unlike the case of normal electron-ion plasma where the pressure follows the ideal gas relation $P=nT$ where $n$ is the number density and $T$ is the temperature, the dust pressure is expected to be complex due to presence of interactions. A number of authors have studied the equation of state of 2D and 3D Yukawa particles both analytically\cite{pandey2004thermodynamics,pandey2005entropy,totsuji2004thermodynamics,avinash2006equation,vaulina2009thermodynamic,avinash2010plasma,avinash2010thermodynamics,oxtoby2013ideal} and using MD simulations\cite{farouki1994thermodynamics,djouder2016equation,Feng_1,feng2016equations,kryuchkov2017thermodynamics}. Most of these simulations have been carried out using periodic boundary conditions. Earlier, Farouki and Hamaguchi carried out the 3D MD simulation for Yukawa particles and showed that while in the weakly coupled phase the dust pressure is  $n_dT_d$, in the fluid phase it becomes strongly negative\cite{farouki1994thermodynamics}. The equation of state of two dimensional monolayer has also been studied recently e.g. Djouder \emph{et al.} have investigated the ideal gas behavior of 2D Yukawa crystal near zero temperature and obtained a relation between pressure and density by fitting the MD simulation results into a power function\cite{djouder2016equation}. Feng \emph{et al.} have established an expression for 2D Yukawa fluids taking MD data from a wide range of parameters varying from melting point to the liquid state\cite{Feng_1,feng2016equations}. 
Kryuchkov \emph{et al.} have studied the thermodynamics of 2D Yukawa systems across coupling regimes; from weakly coupled non ideal gas to strongly coupled fluid and crystalline phase and established a relation between excess energy and excess pressure\cite{kryuchkov2017thermodynamics}. 

In dusty plasma experiments, the dust is usually confined in a cloud within the background plasma by external fields. In this situation there are large mean electric fields present within the dust cloud which contributes appreciably to dust pressure even in the weak coupling regime. This has been confirmed in dusty plasma experiments where dust pressure was measured in weakly coupled dust cloud suspended within the plasma discharge \cite{Fisher2013}. Recently a model which takes into account the confinement of the dust within the plasma has been proposed \cite{avinash2010plasma,avinash2010thermodynamics}. In this model, the dust electrostatic pressure $P_E$, which is the dust pressure due to electrostatic fields present in the cloud, is calculated and shown to scale as $P_E \propto n_d^2$ in the weak coupling regime. For typical parameters of the experiments this contribution is found to be much bigger than the dust thermal pressure $n_dT_d$ consistent with experimental results \cite{quinn2000experimental,williams2007measurement,fisher2010thermal,Fisher2013}. These results indicate the contributions of electrostatic fields present in dusty clouds, even in weak coupling regime is significant and should be taken into account while calculating the dust pressure. Obviously these important contributions which essentially arise due to dust confinements can not be captured in simulations carried out with periodic boundary conditions. In the model of Hamaguchi and Farouki, the dust and plasma both occupy the same volume\cite{hamaguchi1994thermodynamics,farouki1994thermodynamics}. The MD simulation of this model with periodic boundary condition then simulates an infinite dispersion of dust in which dust is confined due to uniform plasma background. In these simulations, thus, the contribution to dust pressure from mean ES fields can not be exemplified. It should be noted that quadratic scaling of dust pressure has been observed earlier in other simulations\cite{charan2014properties,saxena2012dust} and experiments \cite{saitou2012bow}. We defer further discussion on this issue to the last section.  

In this paper we carry out an MD simulation of dusty plasma to critically examine the role of confinement effects on dust pressure. Specifically we obtain an isothermal equation of state (keeping dust temperature constant) of 3D Yukawa gas in the rare density weak coupling regime. 
Based on Avinash's model\cite{avinash2010thermodynamics} we propose following isothermal equation of state for 3D Yukawa gas in rare density weak coupling regime. The total dust pressure $P_d$ at constant dust temperature $T_d$ is given by \:
\begin{equation}
 P_d= n_dT_d + \left( \frac{q_d^2  T}{2 q^2 n}\right) n_d^2 \hspace{0.1cm}, \hspace{0.8cm} T_d=\text{constant},
 \label{Eq:1}
\end{equation}
where $n_d$ and $n$ are the dust and plasma density, and $T$ is the background plasma temperature. It should be noted that isothermal is with respect to dust temperature $T_d$ only. The plasma heat bath always maintains the plasma at constant temperature $T$. The total dust pressure is sum of kinetic pressure $P_K$ of the dust which scales linearly with density and electrostatic pressure $P_E$ which has quadratic scaling with density. In order to take into account confinement effects we consider perfectly reflecting boundary condition in our MD simulations. The confinement effects may also be accounted for by considering a confining potential\cite{saxena2012dust,djouder2016equation} however we have chosen to use reflecting boundary conditions which are simpler to implement. The results of the MD simulations show that in weak coupling regime the electrostatic pressure scales as $ n_d^2$  and gives a significant contribution to the dust pressure . 

Paper is organized in the following manner. In Sec. II we give the details of simulation and the method used for pressure calculation. In Sec. III we present our simulation results and equation of state. In section IV a theoretical account for our results is discussed. In Sec. V we summarize our results and discuss the future scope of work.

\section{Simulation Details}
A large scale Molecular Dynamics simulation is performed using an OpenMP parallel 3 Dimensional Molecular Dynamics (3DMD) code developed by the authors. The further details regarding method, units and equations are given in the following subsections.  
\subsection{Equations and Units}
The  potential energy for $N_d$ number of grains embedded in electron ion plasma is 
 \begin{equation}
 U = \frac{Q_d^2}{4\pi\epsilon_0} \frac{1}{2} \sum_{i=1}^{N_d}\sum_{j,j\neq i}^{N_d}\frac{e^{-r_{ij}/\lambda_D}}{r_{ij}},
 \label{Eq:2}
 \end{equation}
 where $ r_{ij}= |\vec{r_i}-\vec{r_j}|$.  The corresponding force on $i^{th}$ particle due to all other particle is given by 
\begin{equation}
 \vec{F_i}= m \frac{d^2 \vec{r_i}}{dt^2}= \frac{Q_d^2}{4\pi\epsilon_0}\sum_{j,j\neq i}^{N_d}
 \left(1+\frac{r_{ij}}{\lambda_D}\right) \frac{e^{-r_{ij}/\lambda_D}}{r_{ij}^3} \vec{r}_{ij},
 \label{Eq:3}
\end{equation}
 where we have assumed that all grains have equal charge $Q_d$ and mass $m$.

All the lengths have been measured in units of background Debye length $\lambda_D$ and time is measured in units of $\omega_0^{-1}$ (defined later). Therefore, 
$ r \rightarrow \tilde{r}\lambda_D $  and $t \rightarrow \tilde{t} \omega_0^{-1}$, where $\tilde{r}$ and $\tilde{t}$ are dimensionless length and time respectively. Hence Eq. (\ref{Eq:3}) becomes,
$$
 \frac{d^2 \vec{\tilde r_i}}{d\tilde t^2}=\frac{Q_d^2}{4\pi\epsilon_0 m \omega_0^2 \lambda_D^3}\sum_{j,j\neq i}^{N_d} 
(1+ \tilde r_{ij}) \frac{e^{-\tilde r_{ij}}}{ \tilde r_{ij}^3} \vec{ \tilde r}_{ij}.
$$
Let us define $\omega_0$ as $\omega_0 = \left(\frac{Q_d^2}{4\pi\epsilon_0 m  \lambda_D^3} \right)^{1/2}$. This  $\omega_0$  is used for time normalization and integration time steps. The normalized equation of motion for $i^{th}$ particle in normalized units becomes
\begin{equation}
\frac{d^2 \vec{\tilde r}_i}{d\tilde t^2}=\sum_{j,j\neq i}^{N_d}(1+ \tilde r_{ij}) \frac{e^{-\tilde r_{ij}}}{ \tilde r_{ij}^3} \; \vec{ \tilde r}_{ij}.
\label{Eq:4}
\end{equation} 
Eq. (\ref{Eq:4}) is integrated using Leapfrog integrator scheme.

All the energies and temperature ($k_B T$) is measured in units of $Q_d^2/4\pi\epsilon_0\lambda_D$ i.e. $\tilde{E}=E/(\frac{Q_d^2}{4\pi\epsilon_0 \lambda_D})$ and $\tilde{T}_d= k_BT_d/(\frac{Q_d^2}{4\pi\epsilon_0\lambda_D})$.
The dimensionless kinetic per particles can be written as $\tilde{E}_K=({1}/{2N_d}) \sum_{i=1}^{N_d} \tilde{v}_{i}^2$ and 
potential energy as $\tilde{E}_{pot}=({1}/{2 N_d})\sum_i^{N_d} \sum_{j, j\ne i}^{N_d} e^{-\tilde{r}_{ij}}/\tilde{r}_{ij}$.
Temperature is given by mean kinetic energy per particle which, in dimensionless units in 3D, turns out to be
$\tilde{T}_d = ({1}/{3 N_d})\sum_{i=1}^{N_d} \tilde{v}_{i}^2$.
Normalized pressure is defined as, $\tilde{P_d} =P_d/ (\frac{Q_d^2}{4\pi\epsilon_0 \lambda_D^4})$ while number density is given as, $\tilde{n}_d = n_d \lambda_{D}^3 $. Thus, in new units $\kappa $ and $ \Gamma $ becomes
\begin{equation}
 \kappa= \left(\frac{3}{4\pi\tilde{n}_d}\right)^{1/3}, \hspace{0.5cm} \Gamma = \frac{1}{\kappa \;\tilde{T}_d},
 \label{Eq:5}
\end{equation}
where we have used the relation $4 \pi a^3 n_d/3 =1$ to obtain $\kappa$.
\subsection{Pressure calculation}
We have adopted the usual mechanical approach of pressure evaluation which involves ensemble average of the instantaneous or microscopic pressure. For a system of $N$ particles in a volume $V$, the microscopic pressure can be
defined as 
\begin{equation}
 \mathcal{P} = \frac{1}{V}\left(\frac{1}{3}\sum_{i}^Nm_i\vec{v}_i \cdot \vec{v}_i +\frac{1}{3} \sum_i^N \vec{r}_i\cdot \vec{f}_i \right).
 \label{Eq:6}
\end{equation}
 The macroscopic pressure $P$ is obtained by taking either a time or an ensemble average i.e.  $P= \langle \mathcal{P}  \rangle $. In case of pairwise potential, this pressure can be written in Virial form as 
\begin{equation}
 P= nk_BT + \frac{1}{3V}\sum_{i=1}^{N} \sum_{j=1, \neq i}^{N} \vec{r}_{ij} \cdot \vec{f}_{ij}, 
 \label{Eq:7}
\end{equation}
where $n=N/V$ is number density, $ \vec{r}_{ij}$ is the intermolecular vector between a molecular pair and $\vec{f}_{ij}$ is the corresponding intermolecular force. The first term in Eq. (\ref{Eq:7}) is the ideal gas contribution whereas the second term gives the contribution coming from the interactions. 

\begin{figure}[h]
\includegraphics[scale=0.69]{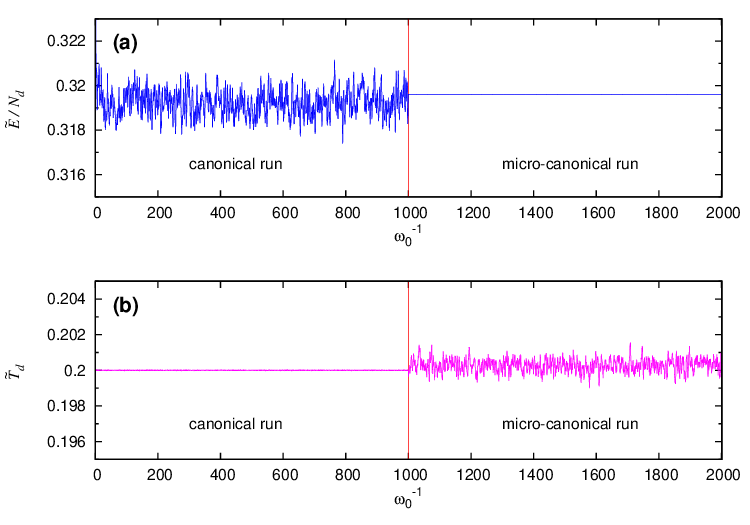}
\caption{\label{fig:1} Temporal evolution of (a) total energy (i.e sum of kinetic and potential energy) and (b) temperature for $\tilde n_d=6.0\times 10^{-3}$ and $\tilde T_d=0.20$. During canonical run the system is coupled with Berendsen heat bath while in micro-canonical run system is kept isolated.}
\end{figure}

\subsection{Method}
To obtain the isothermal equation of state we have varied dust density while keeping the dust temperature constant. Thus, instead of choosing $\Gamma$ and $\kappa$ as free parameters, we have taken $\tilde{n}_d$ and $\tilde{T}_d$ as our input parameters.  The number of particles taken is 4000. We have chosen cubical box with perfectly reflecting boundary conditions. The potential is calculated without any truncation. As the boundary conditions are reflecting, Ewald sum technique is not necessary and has not been used. All the measurements are taken in micro canonical ensemble. To set a desired temperature, we apply Berendsen thermostat for $1000\:\omega_0^{-1}$ and then we let the system go through micro-canonical run for another $1000\:\omega_0^{-1}$. 
Density is changed according to relation $\tilde{n}_d=N_d/\tilde{L}^3$ i.e. by varying the length of cubical box and keeping $N_d$ fixed. The equation of motion given in Eq. (\ref{Eq:4}) is integrated taking time step of size $0.01 \omega_0^{-1}$. As shown in Fig.\ref{fig:1}, this step size is appropriate for conservation of total energy (i.e. sum of kinetic and potential energy).

\section{Results}
Following Eq. (\ref{Eq:1}) we fit our results for pressure according to the following expression,
\begin{equation}
 \tilde{P}_d = \alpha \: \tilde{n}_d + \beta \: \tilde{n}_d^{\;\gamma},
 \label{Eq:8}
\end{equation}
where $\alpha$, $\beta$ and $\gamma$ are constants determined by least square fitting. All the three parameters are treated as free parameters determined simultaneously by minimizing the chi-square. 

\begin{figure*}[]
\begin{center}
 \mbox{\subfigure[][]{\includegraphics[scale=0.71]{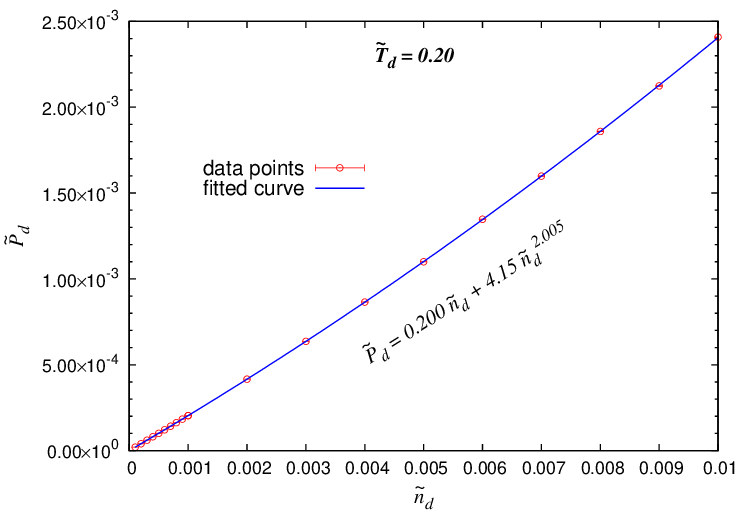} \label{fig:2a}}
   \subfigure[][]{\includegraphics[scale=0.71]{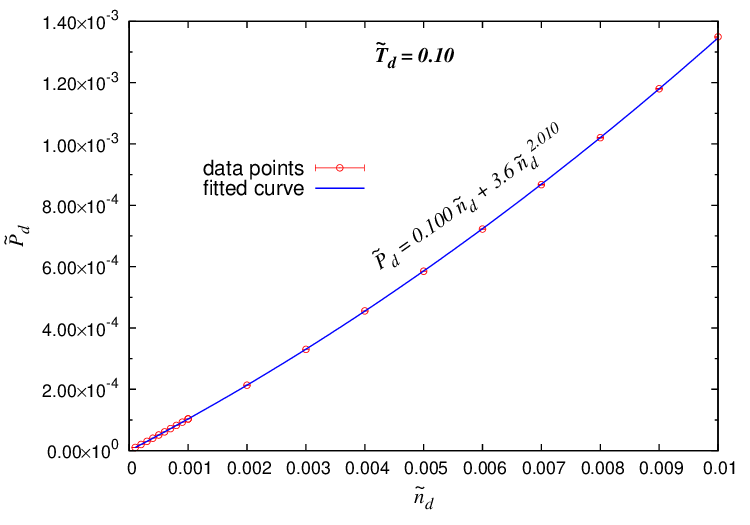} \label{fig:2b} } }
 \mbox{\subfigure[][]{\includegraphics[scale=0.71]{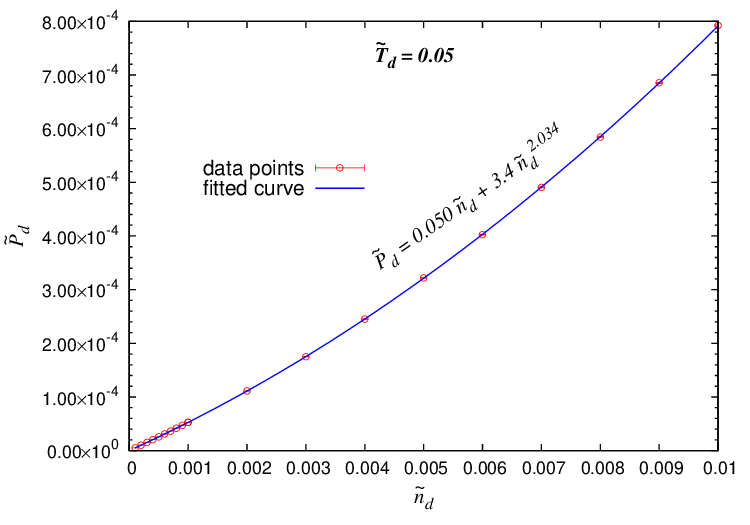} \label{fig:2c}}
   \subfigure[][]{\includegraphics[scale=0.71]{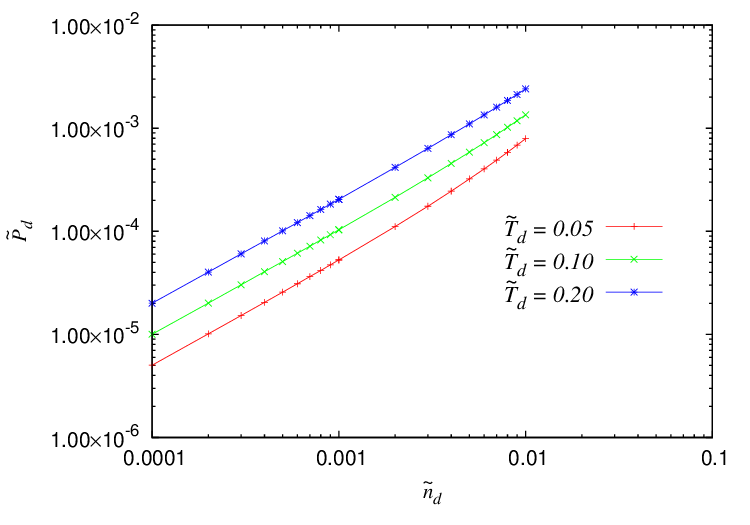}\label {fig:2d}}}  
  \caption{\label{fig:2}  Pressure is plotted versus number density for different dust temperature. Number of particles chosen is $4000$. 
Unit of pressure is  $P/(Q_d^2/4\pi \epsilon_0 \lambda_D^4)$ and unit of density is $n_d\lambda_D^3$.
Fig. (a), (b) and (c)  correspond to isotherms with constant temperature $\tilde{T}_d =0.20$, $\tilde{T}_d =0.10$ and $\tilde{T}_d =0.05$  respectively whereas in Fig. 2(d) these isotherms are shown together on log scale. For a given temperature, dust density is varied from 0.0001 to 0.01. In this regime $\kappa$ changes from $13.365$ to $2.879$ as dust density goes from  $\tilde n_d=1.0 \times 10^{-4}$  to $1.0 \times 10^{-2}$.  
}
 \end{center}
\end{figure*} 

\subsection{Results for $\mathbf{N_d=4000}$}
To examine the dependence of dust pressure on number density, we have chosen three pseudo isotherms corresponding to $\tilde{T}_d =0.20$, $\tilde{T}_d =0.10$ and $\tilde{T}_d =0.05$ respectively. For each dust temperature, dust density is varied from 0.0001 to 0.01. In this regime $\kappa$ varies from $13.365$ to $2.879$ according to relation given in Eq. (\ref{Eq:5}). Thus, this range of $\tilde{n}_d$ is appropriate for examining the low density Yukawa fluid.
\subsubsection{Results for $ \tilde{T}_d=0.20$}
Fig.\ref{fig:2a} shows the results corresponding to $ \tilde{T}_d=0.2$. The fitted parameters are as follows;
$\alpha= 0.2000 $, $\beta=4.1 \: \pm 0.1$ and $\gamma= 2.005 \: \pm 0.008 $.  
The equation for the fitted curve is
$$ \tilde {P}_d = 0.20\; \tilde n_d + 4.1\; \tilde n_d^{\;2.005}. $$ 
The values of $\Gamma^*$ vary from $5.870\times 10^{-7}$ (corresponding to $\tilde n_d = 1.00 \times 10^{-4}$) to $9.753 \times 10^{-2}$ (corresponding to $\tilde n_d = 1.0 \times 10^{-2}$).
\subsubsection{Results for $ \tilde{T}_d=0.10$}
Fig.\ref{fig:2b} shows the results for $ \tilde{T}_d=0.1$. The fitted parameters come out to be $\alpha= 0.1000$ , $\beta=3.6 \: \pm 0.1$ and $\gamma= 2.010 \pm 0.007$ and the fitted equation is
$$ \tilde {P}_d = 0.100\; \tilde n_d + 3.6\; \tilde n_d^{2.010}.$$ 
Here, for $ \tilde{T}_d=0.1$,  $\Gamma^*$ changes from $1.174\times 10^{-6}$  to $0.195$ (corresponding to $\tilde n_d=1.0 \times 10^{-4}$  to $1.0 \times 10^{-2}$).

\subsubsection{Results for $ \tilde{T}_d=0.05$}
The results are shown in Fig. \ref{fig:2c} for $\tilde T_d=0.05$. Here $\alpha =0.0500$ , $\beta=3.41 \pm 0.07$ and $\gamma=2.034\pm 0.004$.
The equation fits in the form:
$$ \tilde {P}_d = 0.050\; \tilde n_d + 3.41\; \tilde n_d^{2.034}.$$
As dust density varies from  $\tilde n_d=1.0 \times 10^{-4}$  to $1.0 \times 10^{-2}$, $\Gamma^*$ changes from $2.348\times 10^{-6}$  to $0.390$.

\begin{figure}[]
 \begin{center}
  \mbox{\includegraphics[scale=0.69]{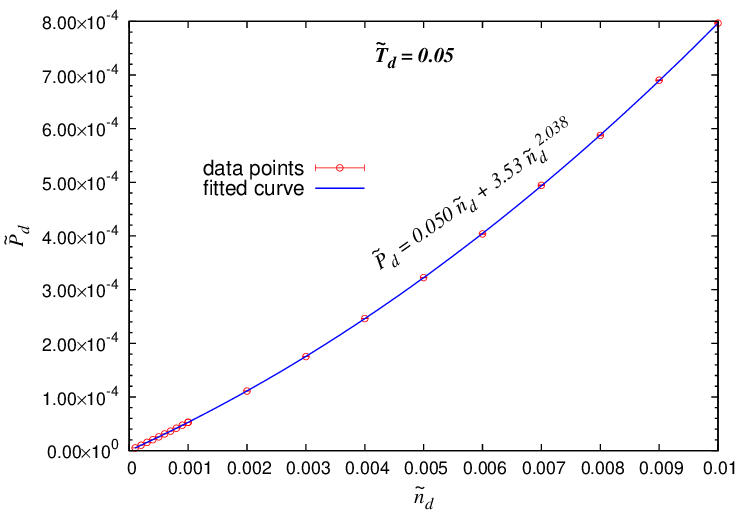} }
   \caption{\label{fig:3}  Pressure versus number density is shown for $N_d=6000$. The isotherm is plotted taking $T_d=0.05$. Comparing the equating of state with that for $N_d=4000$ (fig.\ref{fig:2c}), it is clear that results are almost independent to number of particles chosen. }
 \end{center}
\end{figure}

\subsection{Results for $\mathbf{N_d=6000}$}
To verify that our results are relatively insensitive to number of particles considered in the simulation we have varied the number of particles. Taking $N_d=6000$ we have taken runs for $T_d = 0.05$ to check the scaling of dust pressure with dust density. The range of $\kappa$ and $\Gamma$ is same as for $N_d=4000$. The fitted parameters come out to be $\alpha= 0.0500$ , $\beta = 3.53\pm 0.08$ and $\gamma= 2.038\pm 0.004$ and the equation for this fitted isotherm is
$$ \tilde {P}_d = 0.050\; \tilde n_d + 3.53\; \tilde n_d^{2.038}.$$
As shown in Fig. \ref{fig:3} the scaling for pressure for $N_d=6000$ follows same scaling as for $N_d=4000$.

\subsection{Equation of state}
Based on the results for $\tilde T_d=0.2,\: 0.1$ and $\:0.05$, the following remarks are in order.
For each isotherm, the fitting parameter
\begin{enumerate}[(i)]
 \item $\alpha$ is equal to the temperature i.e. $\alpha=\tilde T_d$, and
 \item $\gamma$ is approximately equal to two i.e. $\gamma \approx 2.0$.
\end{enumerate}
Therefore, the isothermal equation of state for Yukawa gas in rare density regime can be given as
\begin{equation}
 \tilde{P}_d =  \tilde{n}_d \tilde{T}_d + \beta \: \tilde{n}_d^{2.0},
 \label{Eq:9}
\end{equation}
where $\beta$ is a number which ranges from 3.3 to 4.5 for our parameters regimes and depends on the temperature. The reason for temperature dependence of $\beta$ will be discussed in the last section.
 
The first term in the equation of state  corresponds to the usual kinetic pressure whereas the second term  corresponds to ES pressure which depends quadratically on number density. It can be seen from our results that  ES pressure gives a significant contribution to the total pressure of dust.

\section{Discussion of results}
The dust equation of state, in the weak coupling limit, can be analytically obtained from the model of Avinash \cite{avinash2010thermodynamics}. In this model which takes into account the confinement effects, the dust is confined in volume $V_d$ within plasma of volume $V\;(V< V_d)$. Further, the canonical ensemble formulation is used where dust is in contact with heat bath at temperature $T_d$. The Helmholtz free energy of this system is limit given by \cite{avinash2010thermodynamics}  
 \begin{eqnarray}
 F &=& \sum_{\alpha}  T_{\alpha} N_{\alpha} \left(\ln n_{\alpha} \Lambda_{\alpha}^3 - 1\right) \nonumber\\
 &+& \frac{Q_d^2}{8\pi\epsilon_0}  \sum_{i}^{N_d} \sum_{j=1,\neq i}^{N_d} \frac{\exp(-\kappa_D|r_i -r_j|)}{|r_i-r_j|},
 \label{Eq:10}
\end{eqnarray} 
where we have taken ideal gas contribution of electron, ion and dust in first term but still have to take the weak coupling limit in the second term. In this equation $\alpha=(e,i,d)$ i.e. electron, ion and dust, $\Lambda_{\alpha}^3= \left(h^2/2\pi m_{\alpha}k_BT_{\alpha}\right)$, $\kappa_D=1/\lambda_D$, $n_d=N_d/V_d$ while $n_e=N_e/V, \; n_i=N_i/V$. 
The weak coupling limit can be taken in the second term by replacing summation with smooth integration\cite{avinash2010thermodynamics} over dust volume $V_d$ as $\sum_i^{N_d} \sum_j^{N_d} () = n_d N_d \int ()\;d\tau $ to give
\begin{eqnarray*}
\frac{Q_d^2}{8\pi\epsilon_0}  \sum_{i}^{N_d}\sum_{j=1,\neq i}^{N_d} && \hspace{-.5cm} \frac {\exp(-\kappa_D|r_i -r_j|)}{|r_i-r_j|} \\ &=&\frac{Q_d^2}{8\pi\epsilon_0} n_d N_d \int_{V_d} \frac{\exp (-\kappa_D\; r)}{r} d\tau \\
&=& \frac{Q_d^2}{8\pi\epsilon_0} n_d N_d \int_{\bar r} \frac{\exp(-\bar r)}{\bar r \: \kappa_D^2} \: 4 \pi \bar r^2 d\bar r \\
&=&  \frac{ Q_d^2n_d N_d}{2\epsilon_0 \kappa_D^2} X. 
\end{eqnarray*}
where $X \left(=\int_{\bar r} \bar r\: \exp(-\bar r) d\bar{r} \right)$ is a dimensionless number to be fixed by MD simulation. Substituting above results in Eq. (\ref{Eq:10}), we have 
\begin{equation}
F=\sum_{\alpha} T_{\alpha} N_{\alpha} \left(\ln n_{\alpha} \Lambda_{\alpha}^3 -1\right)  +  X \frac{ Q_d^2n_d N_d}{2\epsilon_0 \kappa_D^2}.
\label{Eq:11}
\end{equation}
The dust pressure is given by $P_d=-\left({\partial F}/{\partial V_d}\right)_{T_d}$. Expanding Eq. (\ref{Eq:11}) over $\alpha$ gives
\begin{eqnarray*}
F &=& T_{e} N_{e} \left(\ln n_{e} \Lambda_{e}^3 -1\right) +T_{i} N_{i} \left(\ln n_{i} \Lambda_{i}^3 -1\right) \\
&+& T_{d} N_{d} \left(\ln n_{d} \Lambda_{d}^3 -1\right) + X \frac{ Q_d^2n_d N_d}{2\epsilon_0 \kappa_D^2}.
\end{eqnarray*}
In above equation $V_d$ dependence of $F$ comes from the last two terms via $n_d (=N_d/V_d)$ only. Therefore 
\begin{eqnarray}
P_d &=& - T_{d} N_{d} \left(\frac{\partial}{\partial V_d}\left(\ln n_{d} \Lambda_{d}^3 -1\right)\right)
-X\frac{ Q_d^2N_d }{2\epsilon_0 \kappa_D^2} \left( \frac{\partial}{\partial V_d}n_d\right)   \nonumber \\ 
 &=& n_dT_d + X \frac{ Q_d^2n_d^2 }{2\epsilon_0 \kappa_D^2}.
\label{Eq:12}
\end{eqnarray}
% 
% \begin{equation}
% {P}_d={n}_d {T}_d+X\frac{ Q_d^2\; {n}_d^{2}\; \lambda_D^2}{ \epsilon_0} 
% \label{Eq:12}
% \end{equation}
Above results in MD unit becomes:
\begin{equation}
 \tilde{P}_d =  \tilde{n}_d \tilde{T}_d + 2 \pi X  \: \tilde{n}_d^{2}.
 \label{Eq:13}
\end{equation}
Comparison of Eq. (\ref {Eq:13}) with our simulation results (i.e. Eq. (\ref{Eq:9})) fixes $X$ to $\beta/2 \pi$.  

\section{Summary and conclusion }
To summarize, we have obtained an isothermal equation of state for three dimensional dilute Yukawa gas. We have found by rigorous MD simulations that the pressure for such system not only contains the linear kinetic pressure term but also contains a non linear pressure term proportional to square of the number density. These MD results are consistent with theory and experiments.  We now return to a discussion on periodic boundary condition commonly used in MD simulations vis-\`a-vis perfectly reflecting boundary conditions used in our MD simulations. Let $V$ be the volume occupied by the background plasma and $V_d$ be the volume occupied by the dust. In the model of Hamaguchi and Farouki \cite{farouki1994thermodynamics}, the plasma and the dust occupy the same volume  i.e. $V=V_d$. An MD simulation of this model with Yukawa particles and periodic boundary conditions implies an infinite dispersion of dust which is essentially confined in uniform plasma background and the volume occupied by  dust and uniform plasma is same. In this dispersion the mean electric field is zero and dust pressure is just $n_dT_d$. To see this we consider the Helmholtz free $F$ of Farouki Hamaguchi model\cite{hamaguchi1994thermodynamics}  in the weak coupling limit, including the background plasma contribution given by
\begin{eqnarray}
 F_{HF} &=& \sum_{\alpha} T_{\alpha} N_{\alpha} \left(\ln n_{\alpha} \Lambda_{\alpha}^3 -1\right) 
 -\frac{Q_d^2\,n_d\, N_d}{2 \epsilon_0\,\kappa_D^2} \nonumber \\
&+& \frac{Q_d^2}{8\pi\epsilon_0}  \sum_{i}^{N_d} \sum_{j=1,\neq i}^{N_d} \frac{\exp(-\kappa_D|r_i -r_j|)}{|r_i-r_j|} + PBC,\quad
\label{Eq:14}
\end{eqnarray} 
where $n_{\alpha}=N_{\alpha}/V$ and $PBC$ stands for the Ewald term due to periodic boundary conditions. As discussed in the previous section, the double summation of third term can be approximated as $\sum_i^{N_d} \sum_j^{N_d}  = n_d N_d \int \;d\tau $ in the weak coupling limit which then cancels with the second term and $F_{HF}$ contains only ideal gas contribution in which case $P_d=n_dT_d$. This result as pointed earlier is not consistent with experimental results as the experimentally measured pressure is found to be much bigger than $P_d=n_dT_d$. There are other signs of quadratic scaling of pressure in experiments for example in an experiment on shock formation in a flowing 2D dusty plasma, Saitou \emph{et al.} have shown\cite{saitou2012bow} that the condition of shock formation is satisfied by equation of state $P_d \propto n_d^{\gamma}$ where $\gamma \simeq2.2$. 

It should also be noted that there is evidence that even though the scaling $P_d \propto n_d^{2}$ has been derived with the assumption of weak coupling, it may be very robust; it may  be valid deep in the strongly coupled regime. 
For example, in simulations of Charan \emph{et al.} dusty plasma is confined by external gravity \cite{charan2014properties}. The result of this simulation show $P_d \propto n_d^{2}$ scaling in the bottom region where dust particles are compressed and strongly coupled.
In simulation of Djouder \emph{et al.} dust monolayer is confined using parabolic potential and equation of state is established near zero dust temperature \cite{djouder2016equation}. Their equation of state also suggests $P_d \propto n_d^{2\sim2.165}$ for a wide range of density near dust crystal.  Though $\Gamma^*<1$  for our parameter regime, however the correlation effects are not completely absent in our simulation. The temperature dependence of $\beta$ in our simulation might be attributed to these finite but weak correlations. The dependence of $\beta$ on temperature and the possibility that ES pressure follows the relation $P_d \propto n_d^{2}$  even in the strong couping regime will be addressed in future communication.

\acknowledgments
One of the authors (M.K.S.) acknowledges the financial support from the University Grants Commission (UGC), India, under the SRF scheme. M.K.S. is also thankful to Harish Charan for his kind help regarding the MD simulation.

%\bibliography{Manish_et_al}
%merlin.mbs aipnum4-1.bst 2010-07-25 4.21a (PWD, AO, DPC) hacked
%Control: key (0)
%Control: author (8) initials jnrlst
%Control: editor formatted (1) identically to author
%Control: production of article title (0) allowed
%Control: page (1) range
%Control: year (1) truncated
%Control: production of eprint (0) enabled
%

\end{document}